# Uncertainty-Aware Fuzzy Centrality Measures for Influential Node Identification: A Structural Modeling Approach Toward E-Commerce Applications

Shima Esfandiari
Department of Electrical and Computer Engineering
Shiraz University, Shiraz, Iran

Seyed Mostafa Fakhrahmad*
Department of Electrical and Computer Engineering
Shiraz University, Shiraz, Iran

*Abstract*— In recent years, e-commerce platforms have become one of the most prominent examples of large-scale interaction networks, where understanding influence dynamics among users, products, and digital entities is essential for applications such as online marketing, recommendation systems, and customer behavior analysis. A key challenge in these platforms is that interactions are often uncertain, noisy, and inferred from implicit signals rather than explicitly defined relationships. This uncertainty cannot be effectively captured using deterministic network models. To address this issue, this paper proposes an uncertainty-aware centrality framework for identifying influential entities in e-commerce interaction networks modeled as fuzzy graphs. Unlike existing fuzzy centrality measures that produce single crisp values, the proposed method assigns a fuzzy degree set to each node, preserving multiple possible influence levels along with their associated membership values. Max–min decomposition is employed to enable principled fuzzy inference and accurate uncertainty propagation. Furthermore, a novel ranking strategy is introduced that determines node rankings directly from fuzzy sets without requiring pairwise comparisons, thereby significantly reducing computational complexity. The proposed framework is evaluated on multiple real-world benchmark networks with fuzzy edge annotations, which are widely used as structural proxies for complex interaction systems. Extensive experimental results demonstrate that the proposed method outperforms state-of-the-art centrality measures in terms of ranking accuracy, robustness, and computational efficiency. These findings highlight the effectiveness of the proposed approach as a general methodological tool for influence analysis in e-commerce systems under uncertainty.

## I. INTRODUCTION

In recent decades, e-commerce platforms have experienced rapid growth, becoming complex digital ecosystems where users, products, and services interact at large scale. These platforms can naturally be modeled as complex networks [1], enabling the application of network science techniques to analyze interaction patterns and behavioral dynamics. Through such modeling, it becomes possible to understand and control the spread of information, opinions, and product adoption [2–4], as well as to predict user behavior in response to various online events [5].

One of the most important tasks in the analysis of e-commerce interaction networks is identifying and ranking influential entities [6,7]. Influential users, products, or sellers can significantly affect information diffusion, customer behaviors, and market dynamics [8,9]. Centrality measures play a key role in this context by quantifying the importance of nodes within a network. Well-known centrality measures include degree centrality [10], betweenness centrality (BC) [11], closeness centrality (CC) [12], eigenvector centrality [13], H-index [14], and PageRank [15].

Traditional centrality measures are mainly designed for deterministic networks and have been widely used to identify influential nodes in such settings [16]. Many existing approaches rely on degree-based measures, H-index, or similar metrics to evaluate node importance [16–18]. However, interactions in real-world e-commerce systems are rarely deterministic. Relationships between users, products, and services are often uncertain, ambiguous, or inferred from implicit behavioral data, such as clicks, views, or purchase histories. As a result, deterministic network representations are insufficient to accurately model these systems.

Fuzzy graphs provide a more realistic framework for modeling uncertainty in interaction networks by allowing nodes and edges to have degrees of membership rather than precise binary values. In fuzzy networks, uncertainty can be associated with both nodes and edges. Edge uncertainty may be represented either as a fuzzy set or as a numerical value in the range [0,1]. In this paper, the latter approach is adopted, where an edge weight close to 1 indicates a strong and reliable interaction, while a value near 0 represents a weak or ambiguous relationship.

Several centrality measures have been extended to fuzzy graphs; however, most of these methods ultimately produce crisp numerical values, which leads to a loss of valuable uncertainty information. For instance, fuzzy degree centrality has been defined as the sum of edge weights [19], while fuzzy closeness centrality aggregates the strengths of all paths between nodes [20]. In fuzzy networks, a node may reach another node through multiple paths with different uncertainty levels. In such cases, the strength of a path is determined by the minimum weight of its constituent edges, and among all possible paths, the one with the maximum such value is selected. This inference mechanism is based on the max–min method, which is commonly used in fuzzy logic.

In [18], degree centrality, H-index, and KS centrality are redefined from a fuzzy perspective by assigning a fuzzy set to each node instead of computing a single crisp value. Although this approach preserves uncertainty, the ranking process relies on pairwise comparisons of fuzzy sets, resulting in quadratic time complexity. Building on the same fundamental perspective, this paper proposes a new fuzzy centrality framework that assigns a fuzzy degree set to each node while determining the final ranking directly from the fuzzy set itself. This eliminates the need for pairwise comparisons and significantly improves computational efficiency.

From a methodological standpoint, the proposed framework is designed for influence analysis in e-commerce interaction networks under uncertainty. To evaluate the effectiveness and robustness of the proposed approach, experiments are conducted on several real-world benchmark complex networks with fuzzy edge annotations. These networks exhibit structural properties commonly observed in e-commerce systems, such as heterogeneous degree distributions and sparse connectivity, and are therefore employed as suitable proxies for methodological evaluation. Experimental results demonstrate that the proposed method achieves higher accuracy, greater robustness, and lower computational complexity compared to existing fuzzy centrality measures.

The remainder of this paper is organized as follows. Section II introduces the fundamental preliminaries. Section III reviews related work on centrality measures and fuzzy-based approaches. Section IV presents the proposed uncertainty-aware centrality framework in detail and describes the associated algorithms. Section V evaluates the performance of the proposed method. Finally, Section VI concludes the paper and discusses potential directions for future research in influence analysis for e-commerce platforms.

## II. Preliminaries

### A. Weighted SIR Model

The Susceptible–Infected–Recovered (SIR) model [21] is a widely used diffusion model for simulating influence propagation in networks. It categorizes nodes into three states: susceptible (S), infected (I), and recovered (R). Transitions are controlled by the infection rate β and the recovery rate γ, which is typically set to 1. In the weighted SIR model, the infection rate is proportional to edge weights, allowing uncertainty in interactions to be incorporated into the diffusion process. This model is commonly used for evaluating influence-based centrality measures in fuzzy networks [18].

### B. Fuzzy Graph

A fuzzy graph [22, 23] generalizes classical graphs by assigning membership values in [0,1] to nodes and edges. Formally, a fuzzy graph is defined as G=(V,σ,μ), where V is the set of vertices, σ:V→[0,1] denotes node membership, and μ:V×V→[0,1] represents edge strength. This representation enables effective modeling of uncertainty and ambiguity in real-world interaction networks.

### C. Max–Min Decomposition

Max–min decomposition [23] is a fundamental fuzzy inference mechanism used to evaluate relationships under uncertainty. When multiple paths exist between two nodes, the strength of each path is determined by the minimum edge weight along that path, and the maximum among all such paths is selected. This process is formally defined as:

$$\mu_R(A, B) = \max(\min(\mu_A(x), \mu_B(y))) \quad (1)$$

This mechanism allows uncertainty to be propagated in a principled manner and forms the basis of the proposed fuzzy centrality framework.

## III. Related Work

Identifying influential nodes is a fundamental problem in network analysis and has been extensively studied in various domains, including online social systems and digital platforms related to e-commerce [24-28]. Classical centrality measures such as degree, betweenness, closeness, eigenvector centrality, and PageRank have been widely applied to detect influential entities in deterministic networks. However, these measures assume precise and fully observable interactions, which limits their effectiveness in real-world systems where relationships are often uncertain or imprecise.

To address uncertainty in network interactions, fuzzy set theory, originally introduced by Zadeh [22], has been incorporated into network modeling. Fuzzy-based network representations allow nodes and edges to express partial membership, making them suitable for modeling ambiguous or noisy interactions commonly observed in complex systems. Such models have been applied in diverse application domains, including fault diagnosis, decision-making, and biological and social networks [29,30].

In recent years, considerable attention has been devoted to identifying influential nodes in fuzzy networks. Several approaches have been proposed to extend classical centrality measures by incorporating fuzzy logic. For example, the k-InfNode method integrates fuzzy clustering with random walk mechanisms to identify overlapping and influential nodes. While this approach achieves high accuracy, it is sensitive to parameter tuning and incurs high computational cost in large-scale networks [31]. Similarly, fuzzy-weighted models have been employed to evaluate ambiguous relationships in networks, but they often rely on fixed weights and lack adaptability to dynamic environments [32].

Other studies have introduced fuzzy transfer functions and evolutionary fuzzy optimization techniques to estimate node influence under uncertainty. Although these methods are effective in certain scenarios, they typically require complex parameter configurations and are computationally expensive, which limits their scalability [33,34]. Moreover, some fuzzy-based approaches struggle to handle networks with high levels of overlap and uncertainty propagation.

More recently, fuzzy centrality measures that assign a fuzzy set to each node, rather than producing a single crisp value, have been proposed to better preserve uncertainty information [18]. However, existing methods often rely on pairwise comparisons of fuzzy sets to determine node rankings, resulting in high computational complexity. Motivated by these limitations, this paper proposes an uncertainty-aware centrality framework that preserves fuzzy information while enabling efficient and direct ranking of influential entities. This approach is particularly suitable for influence analysis in large-scale e-commerce interaction networks, where uncertainty and scalability are critical concerns.

## IV. PROPOSED METHOD

This section introduces two uncertainty-aware centrality measures for fuzzy interaction networks: Node Fuzzy Degree Centrality (NFDC) and Node Fuzzy Relationship H-index (NFRH). These measures are designed to identify influential nodes by explicitly modeling uncertainty in edge relationships, which commonly exists in real-world interaction-driven systems.

### A. Node Fuzzy Degree Centrality (NFDC)

In fuzzy networks, edge relationships are uncertain and cannot be represented by crisp values. Therefore, the degree of a node is modeled as a fuzzy variable. Following the max–min decomposition strategy in [18], all possible degree values of a node are enumerated, and a membership degree is assigned to each value based on edge uncertainties. This results in a fuzzy degree set that captures the likelihood of a node attaining different degree levels. Let the fuzzy degree set of node j be defined as $FR(j) = \{(d, \mu_d(j)) \mid d \in (0, DC_j)\}$, where $DC_j$ denotes the crisp degree of node j, and $\mu_d(j)$ represents the membership degree corresponding to degree d

To obtain a final ranking score, the fuzzy degree set is converted into a crisp value as:

$$NFDC(j) = \frac{1}{DC_j} \sum_{d=0}^{DC_j} d \cdot \mu_d(j) \qquad (2)$$

This formulation reflects the expected spreading capability of a node under uncertainty. The detailed computational procedure is summarized in Algorithm 1.

| **Algorithm 1: Computing Node Fuzzy Degree Centrality (NFDC).** |
|---|
| Input: Fuzzy adjacency matrix F, node v |
| 2. Initialize: FR(v) ← ∅ |
| 3. Retrieve all neighbors of v and their edge weights |
| 4. Sort the incident edges of v in descending order of weights |
| 5. for d = 1 to deg(v) do |
| 6.     μ_d(v) ← weight of the d-th edge in the sorted list |
| 7.     FR(v) ← FR(v) ∪ {(d, μ_d(v))} |
| 8. end for |
| 9. Output: Degree fuzzy set FR(v) |

### B. Node Fuzzy Relationship H-index (NFRH)

NFDC primarily captures the local influence of a node. To further incorporate the influence of neighboring nodes, the Node Fuzzy Relationship H-index (NFRH) is proposed. NFRH extends the classical H-index to fuzzy networks by utilizing the NFDC values of neighboring nodes. For each node, the NFDC values of its neighbors are sorted in descending order, and the NFRH value is defined as the largest integer h such that the node has at least h neighbors with NFDC values greater than or equal to h. This measure jointly considers both the quantity and quality of connections under uncertainty. The detailed steps are presented in Algorithm 2.

| **Algorithm 2: Computing Node Fuzzy Relationship H-index (NFRH)** |
|---|
| 1. Input: Fuzzy adjacency matrix F, node v |
| 2. Initialize: h(v) ← 0 |
| 3. Compute degree set D_v = {d1, d2, ..., dn}, where di = Σ_{u∈N(v)} F(i,u) |
| 4. Sort D_v in descending order |
| 5. for i = 1 to n do |
| 6.     if d_i ≥ i then |
| 7.         h(v) ← i |
| 8.     else break |
| 9. end for |
| 10. Output: NFRH of node v: h(v) |

## V. EVALUATION AND RESULTS

This section introduces the datasets used in the experiments and compares the proposed methods with existing approaches. The evaluation is conducted based on three main criteria: robustness, ranking accuracy (imprecision), and computational complexity. The primary objective of this section is to demonstrate the effectiveness of the proposed Novel Fuzzy Degree Centrality (NFDC) and Novel Fuzzy Relationship H- index (NFRH) in identifying influential nodes in uncertain interaction networks, which are used as proxy networks to reflect the structural characteristics of E- commerce systems.

### A. Dataset

To evaluate the effectiveness of the proposed uncertainty-aware centrality measures, eight real-world interaction networks with diverse topological properties were selected as the experimental testbed. These networks range from small-scale structures, such as Karate (34 nodes), to larger infrastructures like Router (5,022 nodes) [35]. The fundamental structural characteristics—including node count (n), edge count (m), average degree (⟨k⟩), clustering coefficient (C), and degree assortativity (r)—are detailed in *TABLE I*.

While these datasets originate from general social and biological domains, they are employed here as structural proxies for e-commerce ecosystems. Utilizing standard benchmark datasets instead of proprietary platform-specific data ensures method reproducibility and avoids biases introduced by domain-dependent preprocessing or opaque filtering often found in commercial datasets. From a network science perspective, these graphs exhibit key topological features intrinsic to social commerce and recommendation systems, including:

*1) Community Structure:* Analogous to customer segments or niche interest groups in e-commerce.

*2) Scale-free Properties*: Reflecting the "rich-get-richer" phenomenon where a few popular products or influencers dominate the market.

*3) Information Diffusion*: Mimicking word-of-mouth marketing and viral trends among consumers. Therefore, these networks allow for a controlled, domain-independent analysis of how the proposed fuzzy centrality measures identify influential nodes (e.g., key opinion leaders or critical products) under structural uncertainty.

To simulate the inherent ambiguity of e-commerce interactions—such as varying levels of consumer trust, review reliability, or interaction intensity—uncertainty is explicitly modeled by fuzzifying the network edges. Specifically, each edge is assigned a membership value derived from a uniform random distribution in the interval [0,1]. This value represents the confidence level or strength of the relationship between nodes. The resulting fuzzy networks are distinguished by the suffix "W" appended to their original names. To ensure robustness and mitigate the stochastic nature of the fuzzification, this process is repeated three times, and all reported results are averages across these instances.

TABLE I. DATASET

| Network | n | m | <k> | <d> | C | r |
|---|---|---|---|---|---|---|
| Dolphins | 62 | 159 | 5.1 | 3.357 | 0.259 | -0.044 |
| Netsci | 379 | 914 | 4.8 | 6.042 | 0.741 | -0.082 |
| Karate | 34 | 78 | 4.6 | 2.4 | 0.571 | -0.476 |
| Email | 1133 | 5451 | 9.6 | 3.606 | 0.220 | 0.078 |
| PolBooks | 105 | 441 | 8.4 | 3.079 | 0.488 | -0.128 |
| Crime | 829 | 1476 | 3.6 | 5.040 | 0.006 | -0.162 |
| Hamster | 2426 | 16631 | 13.7 | 3.589 | 0.538 | 0.047 |
| Router | 5022 | 6258 | 2.5 | 6.449 | 0.033 | -0.138 |

*B. Robustness*

Robustness refers to the ability of a network to maintain its normal functionality when part of it fails due to malfunction, attack, or removal [36]. In the context of interaction networks, robustness reflects how well the system can tolerate the loss of influential nodes while preserving overall connectivity. To evaluate robustness, nodes are sequentially removed according to their importance ranking produced by each centrality method. After removing node i, the remaining level of connectivity is measured by the relative size of the largest connected component (LCC). The robustness metric is defined as:

$$R = \frac{1}{N}\sum_{i=1}^{N}\frac{S_i}{N-1} \quad (3)$$

where N is the total number of nodes and $S_i$ denotes the size of the LCC after node i is removed. A smaller value of $\frac{S_i}{N-1}$ indicates greater damage to the network, implying that the removed node is highly influential.

In the experiments, the most important nodes identified by each method are removed sequentially, and the robustness value R is recalculated after each removal. Fig. 1 illustrates the robustness curves for different centrality measures.

A method that produces a steeper decline and a smaller area under the robustness curve is more effective in identifying influential nodes. As shown in Fig. 1, both proposed measures (NFDC and NFRH) outperform the baseline methods. Among them, NFDC achieves the best overall performance, while NFRH, designed as an enhanced version of FRH, shows significant improvement over its baseline.

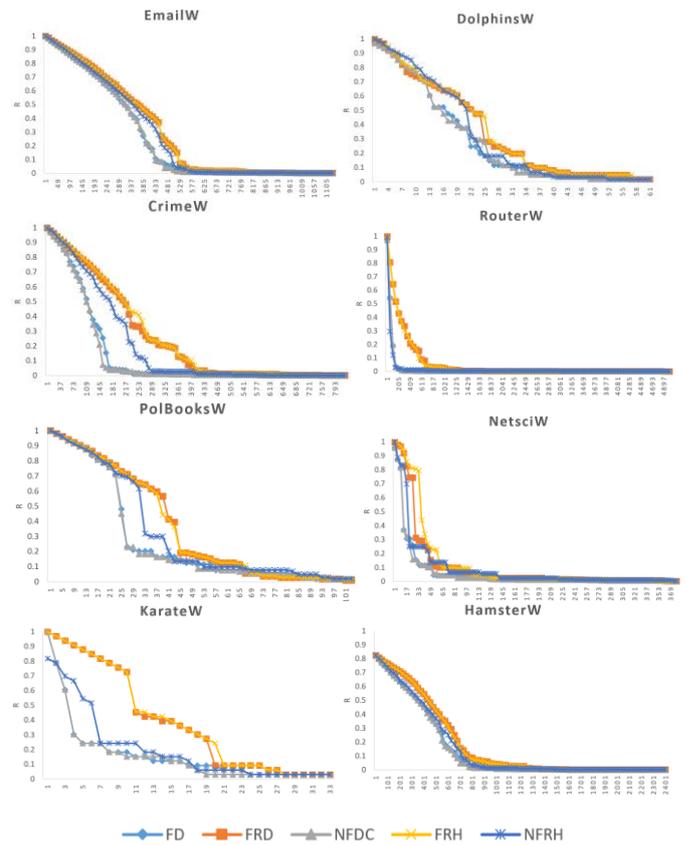

Fig. 1. The robustness of five centrality measures in the dataset.

*C. Imprecision*

To evaluate ranking accuracy, the Imprecision function is employed, as defined in [37]. This metric compares the spreading capability of the top- ranked nodes identified by a given method with their actual spreading effectiveness measured by the Weighted SIR model. The imprecision is defined as:

$$E(P) = 1 - \frac{F_M(P)}{F_{\text{eff}}(P)} \quad (4)$$

where P represents the percentage of top-ranked nodes, $F_M(P)$ denotes the spreading capability of the top P nodes according to ranking method M, and $F_{\text{eff}}(P)$ represents the actual spreading capability obtained from the SIR simulation. Lower values of E(P) indicate higher ranking accuracy.

As illustrated in Fig. 2, the proposed NFDC and NFRH consistently achieve lower imprecision values across all eight fuzzy networks compared to other methods. This confirms their superior ability to accurately identify influential nodes in networks with uncertain and weighted interactions.

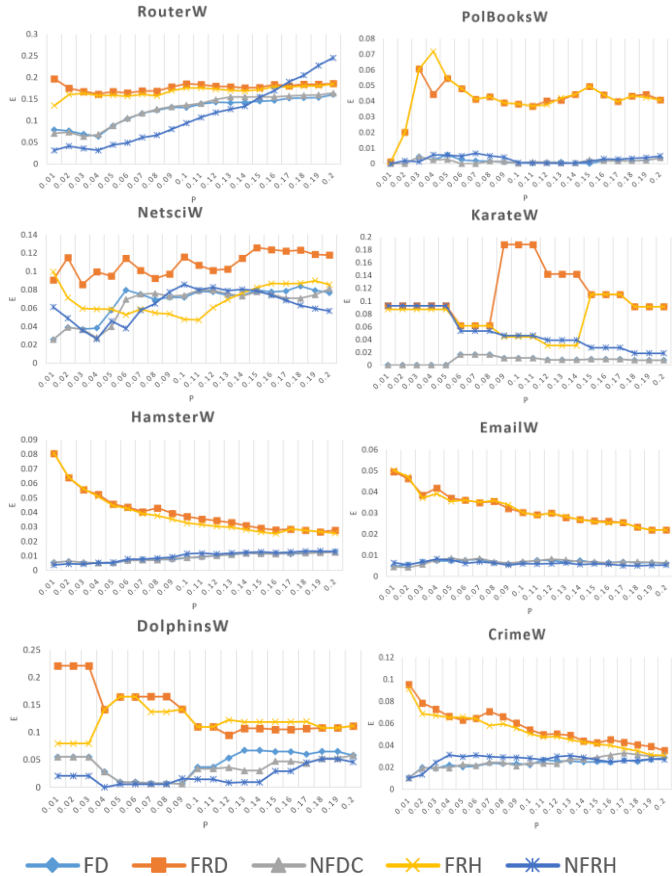

Fig. 2. The imprecision of five centrality measures in the dataset.

### D. Time Complexity

The computational complexity of the benchmark methods and the proposed approach is presented in TABLE II. The complexity of FD is determined by summing the weights of the edges, which corresponds to the number of edges in the network. For NFDC, the computation is slightly more complex than FD because, instead of directly summing the weights, it first sorts them (which has a complexity of $O(\langle d \rangle \log\langle d \rangle)$, where $\langle d \rangle$ represents the average degree in the graph). Then, after constructing the fuzzy set, each degree is multiplied by its corresponding membership degree, resulting in an overall complexity of $O(m+n\langle d \rangle \log\langle d \rangle)$.

In NFRH, the H-index is calculated based on the fuzzy degrees obtained earlier. This also requires sorting the fuzzy degrees, contributing to its computational cost. The FRD and FRH centrality measures have quadratic complexity because they compare the fuzzy sets of each node with all other nodes.

Additionally, Fig. 3 illustrates the execution time of different methods across the dataset networks. It is evident that the FRH and FRD methods require significantly more computational time compared to NFDC and NFRH.

TABLE II. TIME COMPLEXITY OF BENCHMARK METHODS.

| FD | FRD | FRH | NFDC | NFRH |
|---|---|---|---|---|
| $O(m)$ | $O(m(m+1))$ | $O(m(m+\langle d \rangle))$ | $O(m+n\langle d \rangle \log\langle d \rangle)$ | $O(m\langle d \rangle)$ |

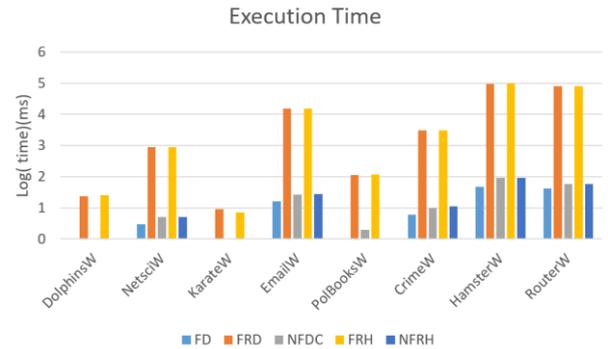

Fig 3. The run time of the five methods in the dataset.

### VI. CONCLUSION

This paper proposed novel uncertainty-aware fuzzy centrality measures designed to identify influential nodes in fuzzy networks, with a specific emphasis on applications in e-commerce scenarios. Recognizing the challenges of accessing real e-commerce data due to privacy and proprietary constraints, we introduced the concept of using structurally representative proxy networks. These public real-world networks capture fundamental features of e-commerce interactions, such as heterogeneous relationships, uncertainty in connections, and hierarchical influence patterns.

Our proposed methods, Node Fuzzy Degree Centrality (NFDC) and Node Fuzzy Rank H-index (NFRH), demonstrated superior performance compared to state-of-the-art approaches in terms of ranking accuracy, robustness against node removals, and computational efficiency. The application of fuzzy membership values to edges effectively models the inherent uncertainty prevalent in e-commerce networks.

This study highlights that structurally chosen proxy datasets provide a reliable and reproducible testbed for evaluating centrality measures under uncertainty, enabling the broader research community to benchmark and compare methods fairly. Future work will focus on extending the framework to dynamic fuzzy networks and exploring integration with real-world e-commerce datasets as they become more accessible, to further validate and refine the proposed centrality metrics.